\documentclass[10pt]{article}
\usepackage{float}
\usepackage[symbol]{footmisc}
\usepackage[superscript,sort]{cite}
\usepackage{array}
\usepackage[normalem]{ulem} %strikeout a word
\usepackage[colorlinks=true,linkcolor=black,citecolor=black,urlcolor=black,bookmarks=true,breaklinks=true]{hyperref}
\usepackage[usenames]{color}
\usepackage{amsmath,amssymb}

\usepackage{amsthm,amscd,amsxtra,amsfonts,amsmath,amssymb,multirow}
\usepackage{wrapfig}
\usepackage[footnotesize]{caption}
\usepackage{subcaption}
\usepackage[tiny,compact]{titlesec}
\usepackage{threeparttable} %add footnote underneath the table
\usepackage{helvet}

\usepackage{xcolor,colortbl}
\usepackage{tikz}
\usetikzlibrary{shapes,arrows}
\usepackage[T1]{fontenc}
\usepackage[linesnumbered,ruled]{algorithm2e} %algorithm
\titlespacing*{\section}{0pt}{*0}{*0}
\titlespacing*{\subsection}{0pt}{*0}{*0}
\titlespacing*{\subsubsection}{0pt}{*0}{*0} 
\titlespacing{\paragraph}{0pt}{*0}{*1}

\definecolor{MyPurple}{rgb}{1,0,1}

\newcommand{\beq}[1]{\begin{equation} \label{#1}}
\newcommand{\eeq}{\end{equation}}
\newcommand{\barray}{\begin{array}{ll}}
\newcommand{\earray}{\end{array}}

\begin{document}
\pagenumbering{roman}

\clearpage \pagebreak \setcounter{page}{1}
\renewcommand{\thepage}{{\arabic{page}}}

\title{Rigidity strengthening is a vital mechanism for protein-ligand binding
}

\author{
Duc Nguyen$^1$,
Tian Xiao$^1$, Menglun Wang$^1$ and
Guo-Wei Wei$^{1,2,3}$ \footnote{ Address correspondences  to Guo-Wei Wei. E-mail:wei@math.msu.edu}\\
%\address{
$^1$ Department of Mathematics \\
Michigan State University, MI 48824, USA\\
$^2$  Department of Biochemistry and Molecular Biology\\
Michigan State University, MI 48824, USA \\
$^3$ Department of Electrical and Computer Engineering \\
Michigan State University, MI 48824, USA \\
}

\date{\today}
\maketitle

\begin{abstract}
Protein-ligand binding  is  essential to almost all life processes. The understanding of protein-ligand interactions is fundamentally important to rational drug design and protein design. Based on  large scale data sets, we show that protein rigidity strengthening or flexibility reduction is a pivoting mechanism in protein-ligand binding. Our approach  based solely on rigidity is able to unveil a surprisingly long range contribution of four residue layers to protein-ligand binding, which has a ramification for drug and protein design. Additionally, the present work reveals that among various pairwise interactions, the short range ones within the distance of the van der Waals diameter are most important. It is found that the present approach outperforms all the other state-of-the-art scoring functions for protein-ligand binding affinity predictions of two  benchmark data sets. 

\end{abstract}
%Key words:
%Protein flexibility,
%Thermal fluctuation,
%Continuum elasticity,
%Atomic rigidity,
%Multiscale.
%PACS numbers: 87.10.-e, 87.14.et,  87.15.Ya, 02.10.Yn
\maketitle
%\section{Introduction}

\paragraph{Introduction}
Protein-ligand binding is fundamental to many biological processes in living organisms. The binding process involves detailed molecular recognition, synergistic protein-ligand corporation, and possible protein conformational change.  Agonist binding alternates receptor  function and trigger a physiological response, such as transmitter-mediated signal transduction, hormone and growth factor regulated metabolic pathways, and stimulus-initiated gene expression, enzyme production, and cell secretion, to name only a few.  The understanding of protein-ligand interactions is essential for drug design and protein design, and has been a central issue in molecular biophysics, structural  biology and medicine. A  common belief is that protein-ligand binding is driven by free energy reduction, which is described in intermolecular forces, such as ionic bonds, hydrogen bonds,  hydrophobic effects, and van der Waals (vdW) interactions \cite{gilson1997statistical, Gilson:2007}. 
% The prevalent view is that electrostatic, van der Waals, hydrophobic and hydrogen bond interactions determine protein-ligand binding affinities. 
However, this view has not directly translated  into accurate binding affinity predictions of large scale binding data sets, despite decades of efforts.
Other potential mechanisms, such as flexibility reduction or rigidity enhancement, have been neglected in the current  modeling and computation. 
 
Current understanding of flexibility with respect to protein-ligand binding is very limited. On the one hand, it is well-known  that  flexibility or plasticity of  proteins as well as ligands facilitates the ligand docking during the binding process  \cite{Zavodszky:2005, Tuffery:2012}. On the other hand, protein-ligand binding reduces the system entropy, which favors the disassociation process. Since protein flexibility is  intuitively associated with conformational entropy, binding induced  flexibility reduction is widely regarded as unfavorable to the protein-ligand binding \cite{Grunberg:2006}. This work offers evidence against this prevalent view. 

Thermodynamically, the protein-ligand binding process is described by the binding affinity, i.e., the change in the Gibbs free energy, which can be expressed in terms of enthalpy and entropy changes at a given temperature. The intricate interplay between  enthalpy and entropy and over-simplified association between flexibility and entropy have made the role of flexibility in protein-ligand binding elusive. Fortunately, the availability of vast amount of affinity data bases \cite{RenxiaoWang:2009Compare} makes it possible to direct test out new hypotheses and reexamine existing theories and computational models. Given the importance of protein-ligand binding to a number of biological fields and disciplines, a wide variety of theoretical and computational approaches have been proposed for the binding affinity prediction. One might classifies these scoring functions into four categories, namely physics based, empirically based, knowledge based and machine learning based ones  \cite{LiuJie:2014}. Physics based models consider vdW and electrostatics interactions between protein and ligand, in addition to  hydrogen bonding and solvation effects \cite{Kollman:2000ACR,Ortiz:1995,Yin:2008}. Empirical or regression methods regard binding affinity as a super position of vdW interaction, hydrogen bonding, desolation, and metal chelation, etc \cite{Zheng:2015LISA,Verkhivker:1995PLP, Eldridge:1997}.  Knowledge-based approaches  make use of available protein-ligand binding databases  to define interaction potentials and scoring functions \cite{PMFScore:1999,DrugScoreCSD:2005,ITScore:2006}. Finally, machine learning strategies take advantages of large scale databases and the progress in statistical regression algorithms \cite{Sarah:2011,Ashtawy:2012,brenner2016predicting} to construct scoring functions  that  outperform other existing binding predictors \cite{BaoWang:2016FFTB}.  However, the winning of machine learning strategies over the physics  based models and the dependence of these strategies on numerous,  sometimes,  apparently unphysical descriptors \cite{BaoWang:2016FFTB} make the molecular mechanism of protein-ligand binding more elusive than ever before.  
 
The objective of the present work is to elucidate the essential role of flexibility or rigidity in protein-ligand binding. We postulate that binding induced protein flexibility reduction, or rigidity strengthening, plays a unique and vital role in the protein-ligand binding. This hypothesis guides us to  design a purely rigidity based machine learning strategy for the prediction of protein-ligand binding affinities.  Protein rigidity modeling is carried out using flexibility-rigidity index (FRI) \cite{KLXia:2013d}.  FRI has been introduced as a simple and efficient algorithm for protein flexibility and  thermal fluctuation analysis \cite{KLXia:2013d}. It has been shown to be about 20\% more accurate and orders of magnitude more efficient than other classic approaches, such as Gaussian network model (GNM) \cite{Bahar:1997}, and anisotropic network model (ANM) \cite{Atilgan:2001} in the B-factor prediction of hundreds of proteins  \cite{KLXia:2013d,Opron:2014,Opron:2015a} and protein-nucleic acid complexes \cite{Opron:2016a}. For example, it predicts the B-factors of the entire HIV capsid with   313,236 residues in less 30 second, which would require GNM more than 120 years to compute were the memory not a problem  \cite{Opron:2014}. Our postulation is confirmed by the evaluation of rigidity index (RI) over 195 protein complexes. Our rigidity index based binding affinity predictor, RI-Score, is able to outperform all other eminent scoring functions, which strongly supports our hypothesis that flexibility reduction or rigidity enhancement is a vital mechanism in protein-ligand binding.  

\begin{figure}[!tb]
	\centering
	\begin{subfigure}[!htb]{0.4\textwidth}
		\centering
		\includegraphics[width=1.0\columnwidth]{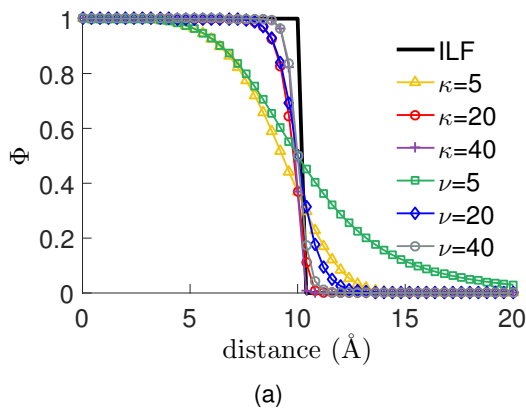}
		\caption{}
	\end{subfigure}
	%\begin{subfigure}[!htb]{0.4\textwidth}
		%\centering
		%\includegraphics[width=1.0\columnwidth]{ilf_2.eps}
		%\caption{}
	%\end{subfigure}
	\caption{Illustration of   FRI correlation functions, which behave like the ideal low filter (ILF)  at large $\kappa$ or $\nu$ values.}
	\label{fig:ilf}
\end{figure}

%\section*{Theory and methods}\label{sec:theory}
\paragraph{Flexibility-rigidity index (FRI)}
 Consider a biomolecule having $N$ atoms with coordinates given as $\{{\bf r_i|r_i}\in\mathbb{R}^3, i=1,2,\dots,N\}$. We denote  $\|\mathbf{r}_i -\mathbf{r}_j\|$ the Euclidean distance between $i$th   and $j$th  atom. 
 We denote $r_i$ the van der Waals radius of $i$th atom and set $\eta_{ij}=\tau(r_i+r_j)$ as a scale to characterize the distance between the $i$th and the $j$th atoms. Where $\tau>0$ is an adjustable parameter. The atomic rigidity index and flexibility index are expressed as
\begin{align}\label{eq:rigidity}
\mu_i=\sum_{\substack{j=1}}^{N} w_{ j} \Phi_{\tau}\left(\|\mathbf{r}_i -\mathbf{r}_j\| \right)  \quad {\rm and}\quad f_i=\frac{1}{\mu_i}, %\label{eq:flexibility}
\end{align}
where $w_j$ are the particle-type dependent weights,  and  $\Phi$ is a real-valued monotonically decreasing correlation function satisfying
\begin{align}\label{eq:admiss}
\Phi \left(\|\mathbf{r}_i - \mathbf{r}_i\| \right)&=1, \quad{\text{as}} \quad  \|\mathbf{r}_i -\mathbf{r}_j\| \rightarrow 0, \\
\Phi \left(\|\mathbf{r}_i - \mathbf{r}_j\| \right)&=0, \quad {\text{as}} \quad  \|\mathbf{r}_i -\mathbf{r}_j\| \rightarrow \infty. 
\end{align}
Commonly used FRI correlation functions   include  generalized exponential functions 
\begin{align}
\Phi^{\rm E}_{\kappa,\tau}\left(\|\mathbf{r}_i -\mathbf{r}_j\|\right)=e^{-\left(\|\mathbf{r}_i -\mathbf{r}_j\|/\eta_{ij}\right)^\kappa}, \quad \kappa>0;\label{eq:expoential}
\end{align}
and generalized Lorentz functions   
\begin{align}
\Phi^{\rm L}_{\nu,\tau}\left(\|\mathbf{r}_i -\mathbf{r}_j\|\right)=\frac{1}{1+\left(\|\mathbf{r}_i -\mathbf{r}_j\|/\eta_{ij}\right)^\nu},\quad \nu>0,\label{eq:lorentz}
\end{align}
As shown in Fig. \ref{fig:ilf},  both generalized exponential functions and  generalized Lorentz functions approximate the ideal low-pass filter (ILF) as $ \kappa \rightarrow \infty $ and $\nu  \rightarrow \infty $, respectively.

	%used in the GNM Kirchhoff matrix \cite{Bahar:1998,Bahar:1997} for B-factor predictions. The ILF function $\Phi(r_{ij}; r_c)$ can be represented as
%\begin{align}\label{eq:ilf}
%\Phi(r_{ij};r_c) = \left\{
%\begin{array}{lr}
%1, & r_{ij}\leq r_c\\
%0, & r_{ij} > r_c.
%\end{array}
%\right.
%\end{align}

  FRI  measures the topological connectivity of the protein-ligand network at every node with appropriate distance-based weights and describes the binding complex with a high level of abstraction. Such an abstraction is ideally suited for the extraction of intrinsic protein-ligand interactions from complex and large-scale binding datasets.  One advantage of FRI is that it allows the multiresolution analysis of protein-ligand binding interactions by varying parameter $\tau$, which endows  FRI  the ability to explore what is the dominant protein-ligand interaction force.    Another advantage of FRI is its multiscale analysis via multi-kernel based multiscale FRI (mFRI) \cite{Opron:2015a}, which enables FRI to capture different length-scales in various protein-ligand interactions.  Finally, by using an ILF representation, the FRI is able to quantitatively detect the relevant length of interactions.

\begin{figure}
	\centering
	\includegraphics[width=0.6\columnwidth]{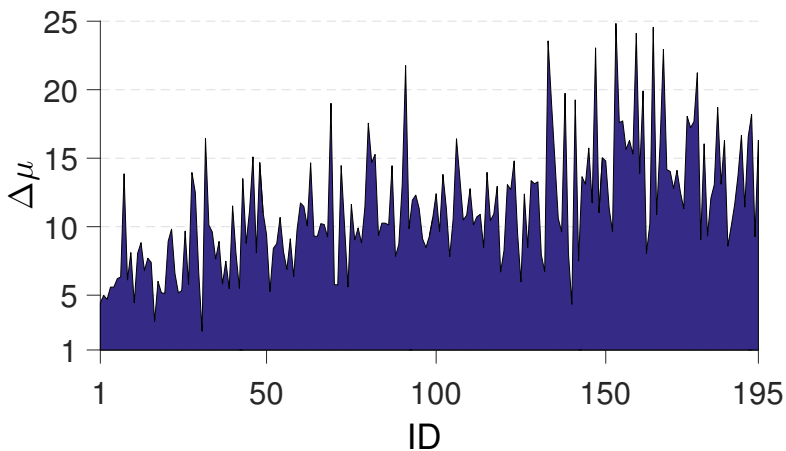}
	\caption{Illustration  relative rigidity change ($\Delta\mu$) upon ligand binding for the PDBbind v2007 core set $(N=195)$.  The positive  relative rigidity changes for all proteins confirms the hypothesis that ligand binding strengthens protein rigidity.
	}
	\label{fig:rigidity}
\end{figure}
      
\paragraph*{Binding induced rigidity strengthening  }
We first  illustrate how the protein rigidity is quantitatively strengthened after ligand binding. To this end, we define  protein relative rigidity change $\Delta \mu ({\rm Com})$  by $\Delta \mu ({\rm Com}) =\sum_{j}\left[\frac{\mu_j({\rm Com})-\mu_j({\rm Pro})}{\mu_j({\rm Pro})}\right]$, where $\mu_j({\rm Com})$ and $\mu_j({\rm Pro})$ are the rigidity indexes  of the $j$th heavy atom of the protein-ligand complex (${\rm Com}$) and the original protein (${\rm Pro}$), respectively\cite{KLXia:2013d}. Figure \ref{fig:rigidity} depicts the protein relative rigidity change  ($\Delta\mu$)   against each complex in the PDBBind v2007 core set (N=195)  \cite{RenxiaoWang:2009Compare}.  It is interesting to note that all  the relative  rigidity changes are positive, which confirms our hypothesis that ligand binding enhances protein rigidity and reduces protein flexibility. 

\paragraph*{Rigidity index based scoring functions (RI-Score)}
Having confirmed that ligand binding enhances protein rigidity, it remains to establish that rigidity strengthening is a vital mechanism for protein-ligand binding. Our hypothesis is that,  for blind prediction of protein-ligand binding affinities, if a quantitative model solely based on rigidity analysis is very accurate and more competitive than other state of the art models in the field, it proves that {\it rigidity strengthening is a vital mechanism for protein-ligand binding}. To prove our hypothesis, we   exclude electrostatic, van der Waals, hydrogen bond, hydrophobic and hydrophilic interactions used in conventional force field based models \cite{Gilson:2007,Kollman:2000ACR,Ortiz:1995,Yin:2008}, including ours \cite{BaoWang:2016FFTB} and  consider nothing but protein rigidity change upon ligand binding. We define  element-specific protein-ligand rigidity index by collecting  cross correlations    
\begin{align}\label{eqn:features}
{\rm RI}^{\alpha}_{\beta,\tau,c} ({\rm X-Y})= \sum_{k\in {\rm X}\in{\rm Pro}}\sum_{l\in{\rm Y}\in {\rm Lig}}  \Phi^{\alpha}_{\beta,\tau}\left(\|\mathbf{r}_k -\mathbf{r}_l\|\right), \ \forall \|\mathbf{r}_k -\mathbf{r}_l\|\le c,
\end{align}
where   $\alpha={\rm E, L}$ is a kernel index indicating either the exponential kernel (${\rm E}$) or Lorentz kernel (${\rm L}$). Correspondingly,  $\beta$ is kernel order index such that $\beta=\kappa$ when $\alpha={\rm E}$ and $\beta=\nu$ when $\alpha={\rm L}$.  We adopt our fast FRI (fFRI) based on the cell lists algorithm \cite{Allen:1987} with a cutoff distance $c$ to reduce computational complexity \cite{Opron:2014}. Here,  ${\rm X}$ denotes a type of heavy atoms in the protein (${\rm Pro}$) and  ${\rm Y}$ denotes a type of heavy atoms in the ligand (${\rm Lig}$).  Four types of protein heavy atoms, namely $\{{\rm C, N, O, S}\}$, and nine types of ligand atoms,  i.e., $\{{\rm C, N, O, S, P, F, Cl, Br, I}\}$, are utilized in this work. Unlike force field based methods which require sophisticated data processing, the current model  applies directly to x-ray crystallography data. 

\section*{Results and discussion}
\paragraph{The PDBBind v2007 benchmark}
For quantitative prediction of protein-ligand binding affinities, we combine protein-ligand rigidity index in Eq. (\ref{eqn:features}) and machine learning to construct RI-Score.  Although machine learning can be a powerful approach for  modeling  massive datasets, its performance depends crucially on its feature vectors. Therefore, if rigidity strengthening is truly the dominant mechanism for protein-ligand binding, the proposed RI-Score should be able to do well in the binding affinity prediction of massive experimental data sets. Although a specific machine learning algorithm, Random Forest, is used in this work, other machine learning techniques, such as gradient boosting algorithms, deliver similar results. To validate RI-Score, we consider a benchmark  dataset, \href{http://www.pdbbind.org.cn/}{PDBBind} v2007, to validate our RI-Score against  a large number of eminent scoring functions  \cite{RenxiaoWang:2009Compare,IDScore:2013,HLi:2015}. More specifically, the PDBBind v2007 core set of 195 protein-ligand complexes is utilized as our test set, while our model is trained on the  PDBBind v2007 refined set of 1300  protein-ligand complexes, excluding the PDBBind v2007 core set of 195  complexes  \cite{RenxiaoWang:2009Compare} (i.e., $N=1105$).  
   
%The
\begin{figure}[!tb]
	\centering
	\begin{subfigure}[!htb]{0.45\textwidth}
		\centering
		\includegraphics[width=1.0\columnwidth]{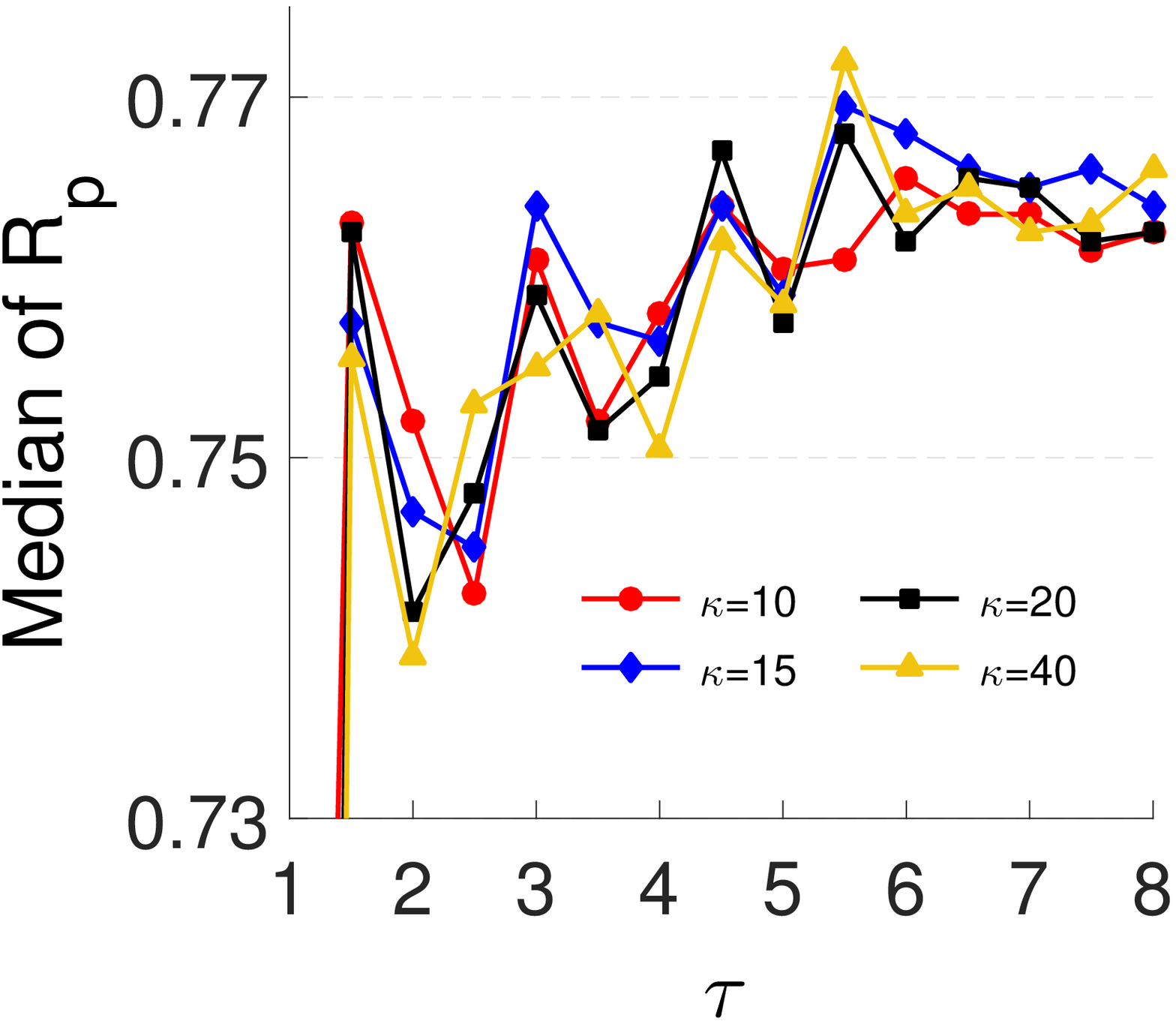}
		\caption{}
	\end{subfigure}
	\begin{subfigure}[!htb]{0.45\textwidth}
		\centering
		\includegraphics[width=1.0\columnwidth]{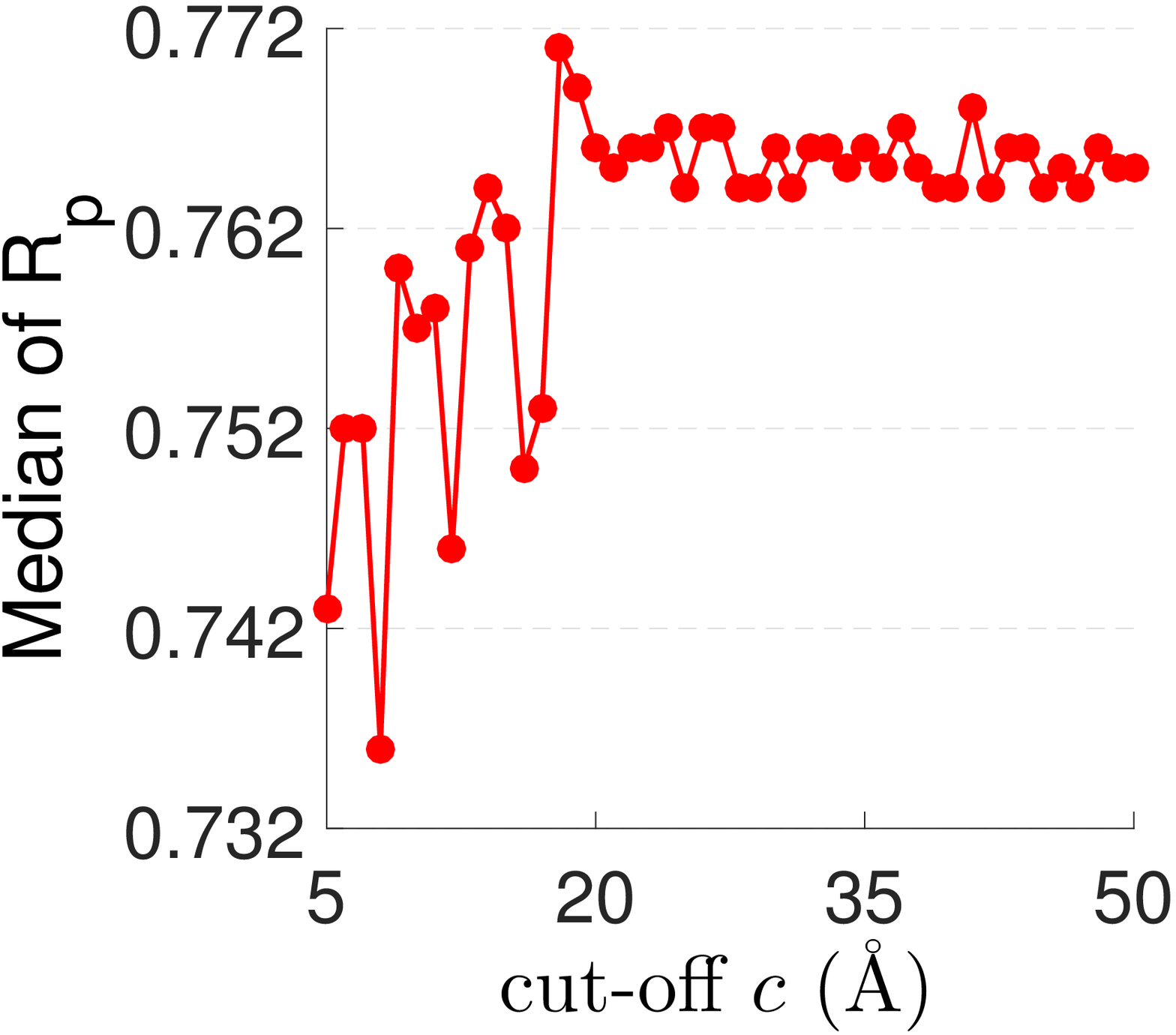}
		\caption{}
	\end{subfigure}
	\caption{(a) Pearson correlation coefficients ($R_p$) of RI$^{\rm E}_{\kappa,\tau,18}$ are plotted against the choice of $\tau$ for for PDBBind v2007 core set over a range of values for $\kappa$. Values of $R_p$ under 0.73 are not shown. (b)  $R_p$ values of RI$^{\rm E}{40,5.5,c}$  are plotted against different values of cut-off distance $c$ for PDBBind v2007 core set. 
	The obvious oscillations are associated with the numbers of nearest residues included in the rigidity features.   
	}
	\label{fig:ilf_p}
\end{figure}

We first consider exponential kernels which are known for their fast decay and thus can be used to detect the effective length of of short-,  medium- and long-range interactions. To this end, we investigate the behavior of high-order exponential kernels with large $\kappa$ values, which are essentially the ideal low pass filter  as shown in Fig. \ref{fig:ilf} and thus are able to exclude all interactions beyond the kernel length scale. 	Figure \ref{fig:ilf_p}(a)  depicts the Pearson correlation coefficients of four  high order exponential kernels over a wide range of scales. It is interesting to find surprising  oscillations in the Pearson correlation coefficients over different scales. Such oscillations indicate the inclusion of different numbers of nearest residue layers. For example, when $\tau=1.5$, $\eta_{\rm CC}$ for a ${\rm C-C}$ pair is 5.1\AA. Since one of these two carbons belongs to the protein and the other belongs to the ligand, the only hydrophobic interactions that can be accounted are those from the nearest layer of residues.   The next peak occurs at $\tau=3$ (i.e.,  $\eta_{\rm CC}=10.2$\AA), which is due to the inclusion of the interactions of the nearest two layers of residues. It is amazing to note that these oscillations persist for two more peaks to reach the best predictions at  $\tau=5.5$, which suggests that  interactions from the nearest four layers of residues effectively contribute to the protein-ligand binding. To our knowledge, it is very difficult for physics based methods to track such  long-range interactions.  The present finding has significant ramifications to protein design.

We  note that when   $\tau=1.5$,   Pearson correlation coefficients obtained by all exponential kernels are already   better than those attained by most other methods in the field \cite{RenxiaoWang:2009Compare,IDScore:2013,HLi:2015}. This good performance  is due to the fact that at $1.5(r_i+r_j)$, all the hydrogen bonds,  most van der Waals interactions, and   good portion  of electrostatic interactions are included in our RI-Score. It is well known that   hydrogen bond interactions can be effective from 2.2 to 4 \AA, with  donor-acceptor distances of 2.2-2.5\AA~ as strong, mostly covalent, 2.5-3.2\AA~ as moderate, mostly electrostatic, and 3.2-4.0\AA~ as weak, electrostatic  \cite{Jeffrey:1997}. Their corresponding energies are about 40-14, 15-4, and less than 4 kcal/mol, respectively \cite{Jeffrey:1997}.

The impact of cutoff distance as shown in 	Figure \ref{fig:ilf_p}(b) also show an oscillatory pattern. In fact, this oscillation is consistent with that shown in 	Figure \ref{fig:ilf_p}(a). 

\begin{figure}[!tb]
	\centering
	\begin{subfigure}[!htb]{0.45\textwidth}
		\centering
		\includegraphics[width=1.0\columnwidth]{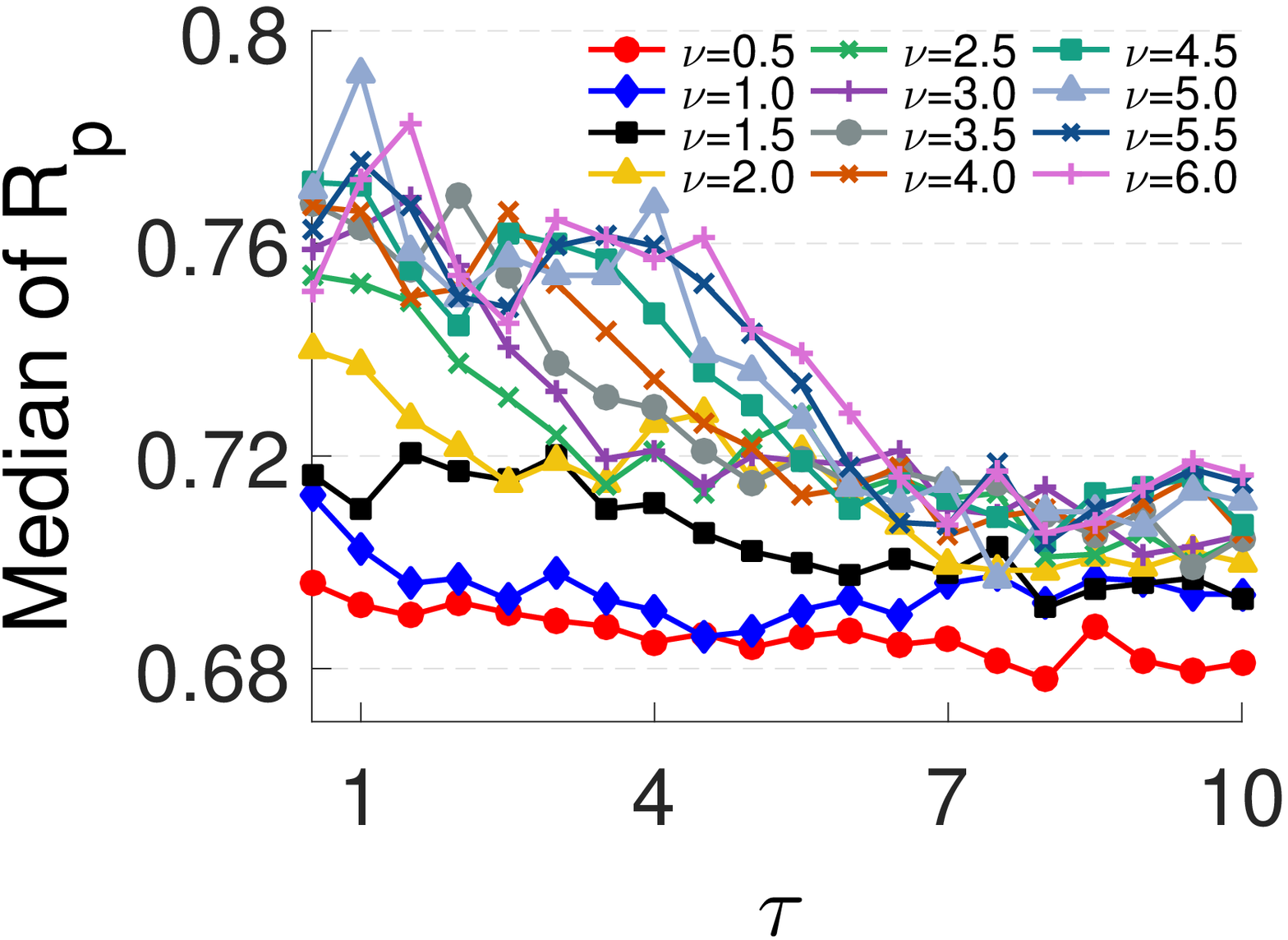}
		\caption{}
	\end{subfigure}
	\begin{subfigure}[!htb]{0.45\textwidth}
		\centering
		\includegraphics[width=1.0\columnwidth]{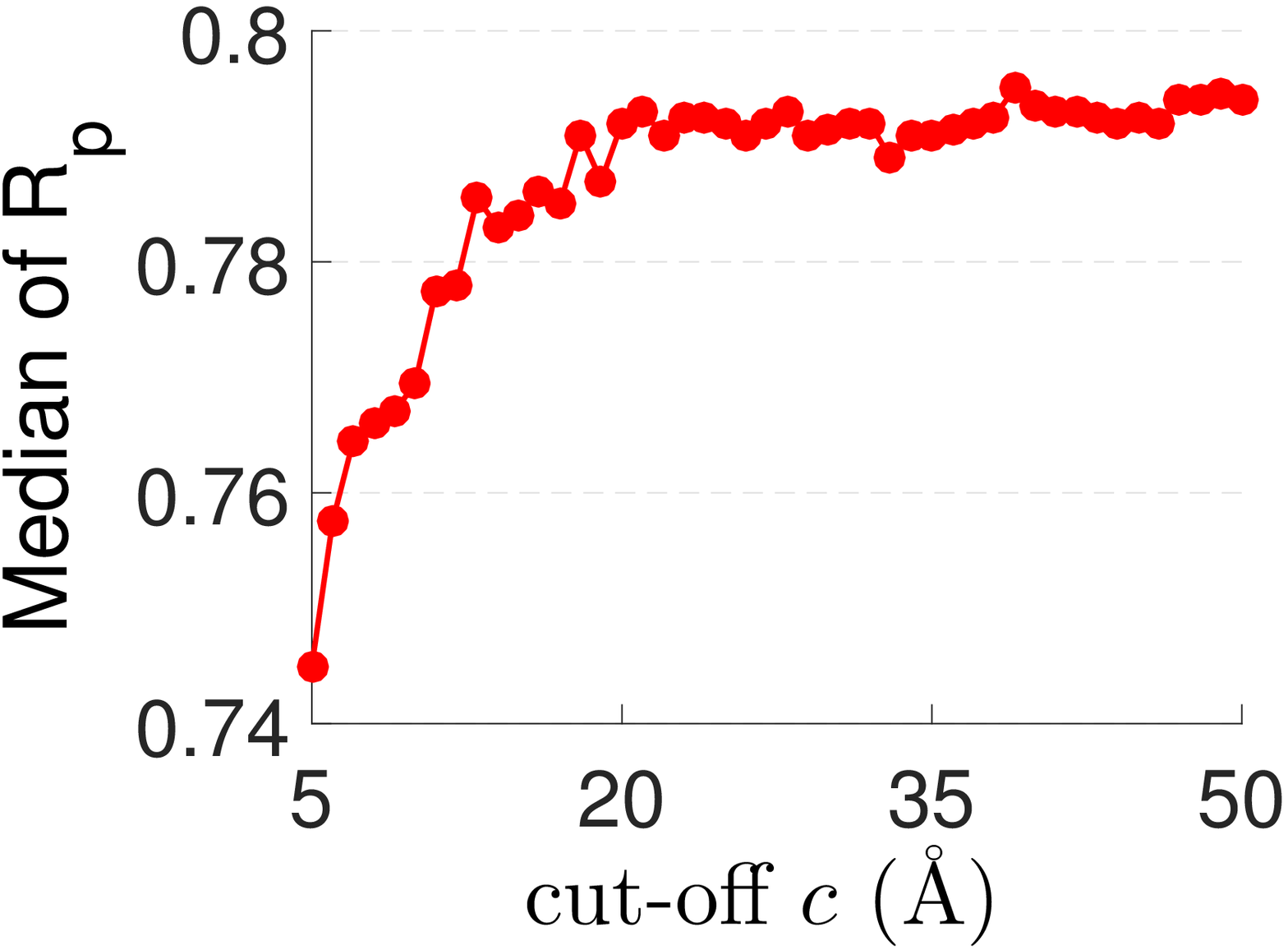}
		\caption{}
	\end{subfigure}
	\caption{(a) Pearson correlation coefficients ($R_p$) of RI$^{\rm L}_{\nu,\tau,40}$ are plotted against the choice of $\tau$ for for PDBBind v2007 core set over a range of  $\nu$ values.
	 (b) $R_p$ values of RI$^{\rm L}_{5,1,c}$ for PDBBind v2007 core set are plotted against different values of cut-off distance $c$.}
	\label{fig:lor_p}
\end{figure}

Unlike exponential kernels, Lorentz kernels are well known for their slow decay, which is able to capture interactions over a wide range of distances. Figure \ref{fig:lor_p} shows the influence of the power of the Lorentz kernel, the scale, and cutoff distance to the blind prediction accuracy in terms of  Pearson correlation coefficients with respect to the experimental binding affinity data.  With a sufficiently large cutoff values $c=40$\AA, best prediction is obtained at $\nu=5$ and $\tau=1$. This result reveals that most important protein-ligand interactions occur approximately at  van der Waals distances, which strongly indicates that short range hydrogen bond interactions are relatively more important than long range interactions. When  $\nu=5$ and $\tau=1$, a cutoff value of 20\AA, which indeed contains four layers of residues, is found to be large enough to include all the essential protein-ligand interactions.

\begin{figure}[!tb]
	\centering
	\begin{subfigure}[!htb]{0.45\textwidth}
		\centering
		\includegraphics[width=1.0\columnwidth]{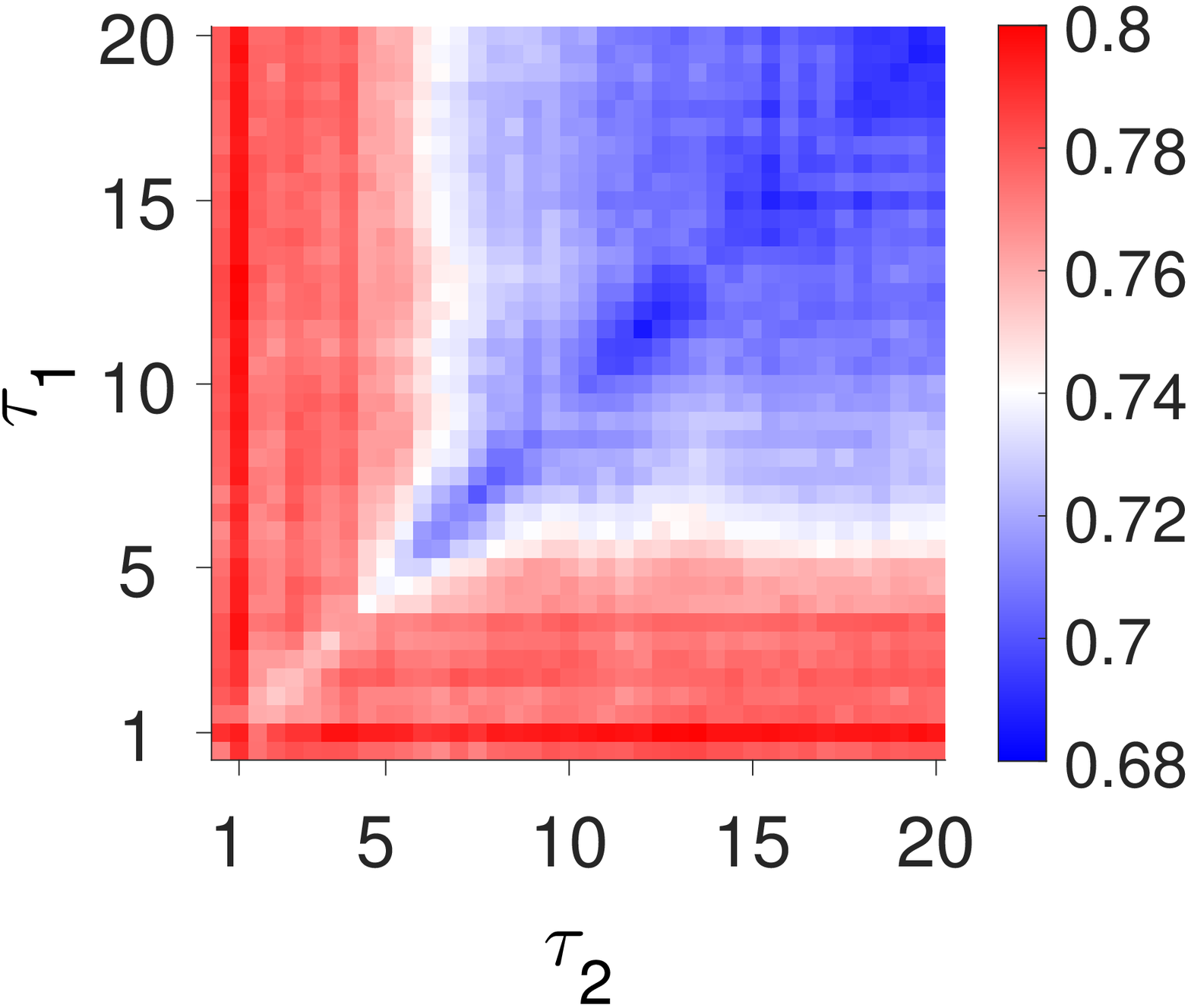}
		\caption{}
	\end{subfigure}\quad
	\begin{subfigure}[!htb]{0.45\textwidth}
		\centering
		\includegraphics[width=1.0\columnwidth]{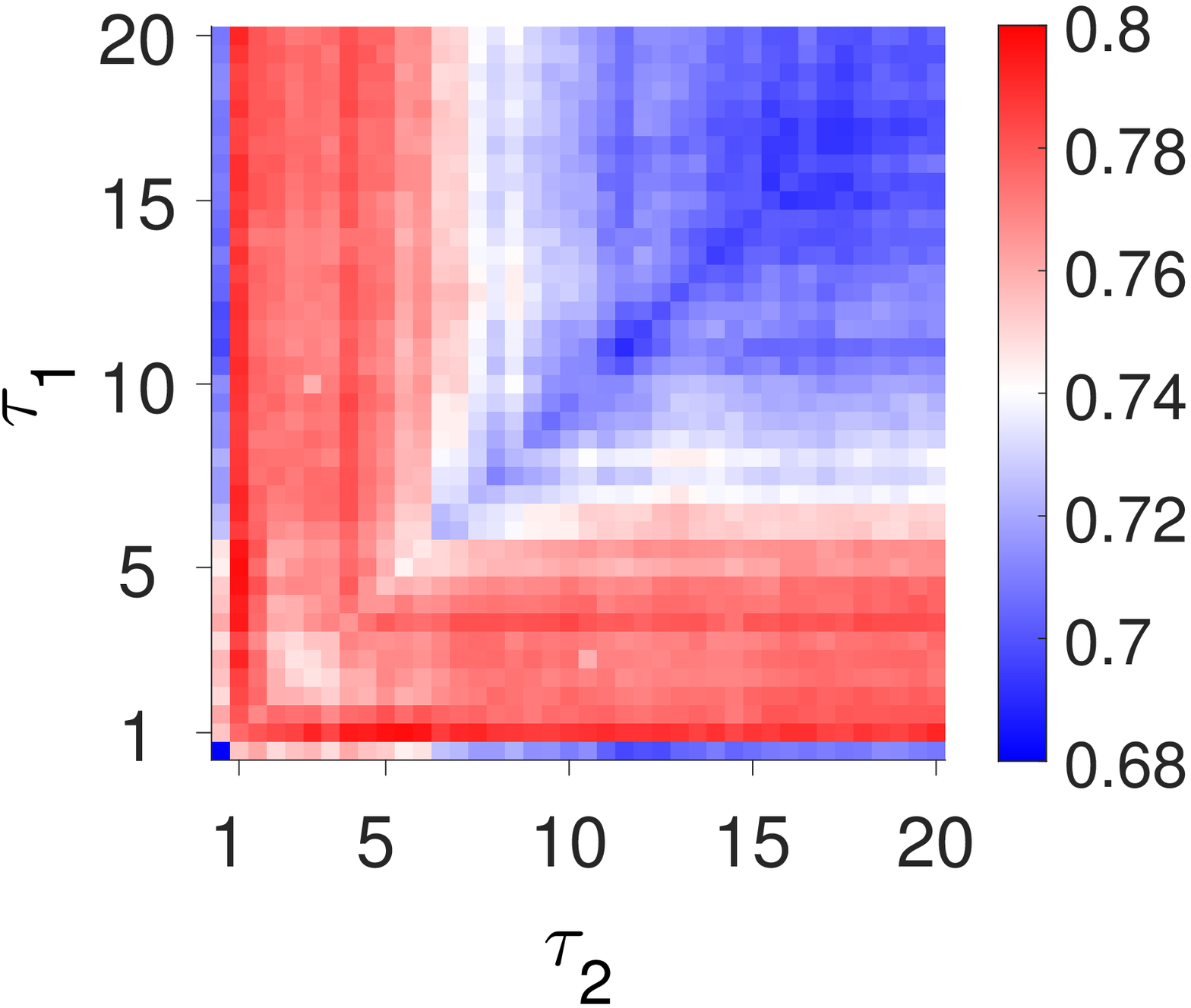}
		\caption{}
	\end{subfigure}
	\caption{An illustration of multiscale behavior in protein-ligand binding prediction. Mean values of Pearson correlation coefficients of 2-scale models on PDBBind v2007 core set are plotted  against different scale values $\tau_1$ and $\tau_2$: (a) RI$^{\rm LL}_{5,\tau_1,\max(39,3.7\tau_1);5,\tau_2,\max(39,3.7\tau_2)}$; (b)  RI$^{\rm EE}_{2.5,\tau_1,\max(12,3.7\tau_1);2.5,\tau_2,\max(12,3.7\tau_2)}$.  Scale parameters $\tau_1$ and $\tau_2$ vary from 0.5 to 20 with increments of 0.5.   Obvious,  the  combination  of a  relatively  small-scale  kernel  and  a  relatively  large-scale  kernel  delivers best prediction, which indicates the importance of incorporating multiscale in protein-ligand binding prediction.}
	\label{fig:2k}
\end{figure}

Protein-ligand interactions  intrinsically  involve multiple characteristic length scales, such as those for three types of hydrogen bonds, van der Waals bonds, and hydrophobic interactions, which can still be very important for residues far away as indicated above. In our earlier study, we found that multiscale FRI (mFRI) that utilizes two or three FRI kernels parametrized with different length scales can significantly improve the B-factors prediction. In this work, we are interested in  examining the impact of multiscale rigidity to binding free energy prediction. Machine learning approach makes it particularly convenient to incorporate multiscale effect in predictive models. One just needs to construct one additional set of features at a desirable scale.  To this end, we construct two sets of RI feature vectors, with one of the set parametrized at $\tau_1$ and the other at $\tau_2$. In general, we denote these feature vectors by  RI$^{\alpha_1\alpha_2}_{\beta_1,\tau_1,c_1;\beta_2,\tau_2,c_2}$ as a straightforward extension of our notation. To reduce the number of degrees of freedom, values of $\nu (\text{or~}\kappa)$ and $c$ are adopted from the corresponding single-kernel model. As a result, we only need to search for two scale parameters $\tau_1$ and $\tau_2$. Figure \ref{fig:2k} illustrates the impact of two-scale RI feature based RI-Score predictions of the PDBBind v2007 core set of 195 complexes. The median of $R_p$ of RI$^{LL}$ and  RI$^{\rm EE}$ are plotted against  different pairs of scale parameters $(\tau_1,\tau_2)$ varying from (0.5,0.5) to (20,20).   First, it is easy to see that  two-scale predictions are typically better than the single scale  one.  The  best  predictions  are typically achieved at  the  combination of a relatively small-scale kernel and a relatively large-scale kernel. A scale optimized model,  RI$^{\rm LL}_{5,1,39;5,12.5,47}$, delivers the  median and the best Pearson correlation coefficient $(R^m_p,R^b_p)=(0.800,0.811)$.  Similarly a scale optimized model,  RI$^{\rm EE}_{2.5,1,12;2.5,5.5,21}$, achieves $(R^m_p,R^b_p)=(0.797,0.809)$. More details  are reported in Table SI in Supporting Information. These findings again confirm the    importance of incorporating multiscale  in  the  protein-ligand predictions.

 %
%
%\begin{figure}
	%\centering
	%\includegraphics[width=0.5\columnwidth]{rmse_2007.eps}
	%\caption{The comparison between experimental binding free energies and RI$^{LLE}$ prediction on PDBbind v2007 core set $(N=195)$. The RMSE and Pearson correlation coefficient are 1.97 kcal/mol and 0.822, respectively. }
	%\label{fig:rmse_2007}
%\end{figure}

\begin{figure}
	\centering
	\includegraphics[width=0.8\columnwidth]{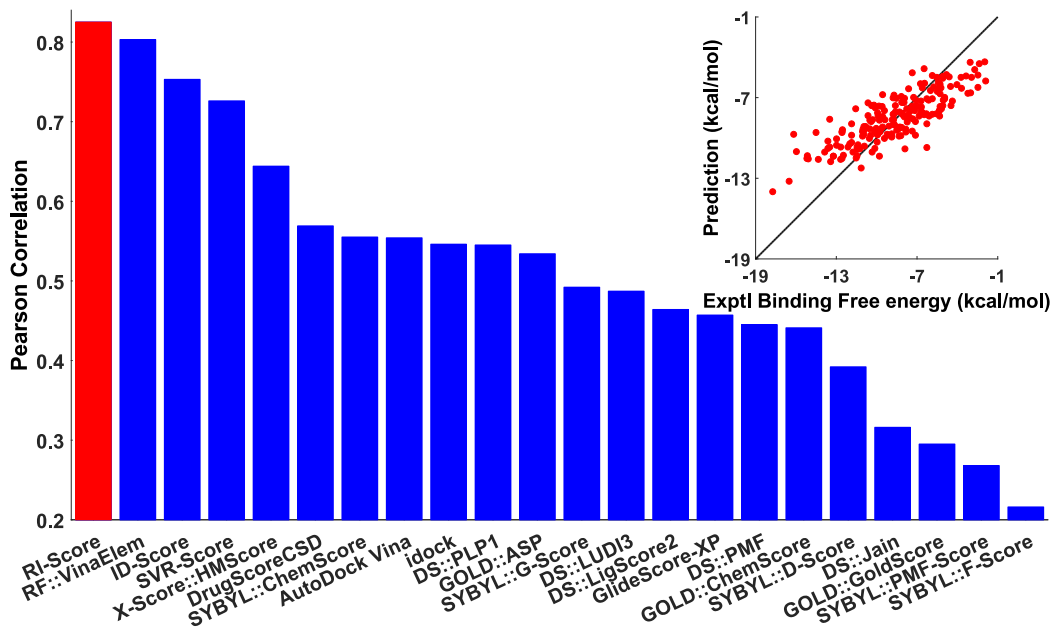}
	\caption{Performance comparison between different scoring functions on the PDBBind v2007 core set. The performances of other scoring functions are adopted from Refs. \cite{RenxiaoWang:2009Compare,IDScore:2013,istar:2014,HLi:2015}. The RMSE and Pearson correlation coefficient of our prediction are 1.99 kcal/mol and 0.825, respectively.}
	\label{fig:v2007_scores}
\end{figure}

Encouraged by the above two-scale results, we have also explored the utility of three-scale RI-Score models the prediction of the PDBBind v2007 core set. Certainly, a complete search of the parameter space is  prohibitively expensive. We therefore focus on exploring the scale parameters $\tau_1$ and $\tau_2$ for certain sets of other parameters. It is found that  RI$^{\rm LLE}$ is the best model   with $(R^m_p,R^b_p)=(0.803,0.825)$. The corresponding RMSE of our best prediction is 1.99 kcal/mol. More details of predicted energies by RI-Score are provided in the Supporting Information.

Predictions from different scoring functions on PDBBind v2007 core set have been presented in the literature \cite{RenxiaoWang:2009Compare,IDScore:2013,HLi:2015}. Figure \ref{fig:v2007_scores} plots the performance of these scoring functions along with our prediction, i.e., RI$^{\rm LLE}$ (highlighted with red color). Clearly, RI-Score outperforms all the other eminent scoring functions.   The correlation between experimental binding free energies and the best prediction attained by RI$^{\rm LLE}$ is also presented in the figure. 

\paragraph{The PDBBind v2013 benchmark}
It remains to show that the outstanding performance of proposed RI-Score is not limited a specific data set. To this end, we consider  PDBBind v2013 core set of 195 protein-ligand complexes as our test set \cite{YLi:2014}. The  PDBBind v2015 refined set of 3706 protein-ligand complexes, excluding the PDBBind v2013 core set, is employed as our training set. Our best model for the PDBBind v2013 core set which is the same as  PDBBind v2015 core set, is given  by  RI$^{\rm LLE}_{6,4,15;6,10,37;3.5,1.5,8}$ with a Pearson correlation coefficient and RMSE of 0.782 and  2.051 kcal/mol, respectively. More details of  RI-Score predictions are provided in Table S3  of the Supporting Information. 

 PDBBind v2013 core set is also a popular test set for benchmarking scoring functions. For comparison, Fig. S8 plots the performance of our method and 25 other scoring functions on this benchmark with data taken from Refs. \cite{YLi:2014}. Again, RI-Score is found to be significantly more accurate than other eminent methods. 

\section*{Conclusion}
Protein-ligand binding is of paramount importance to biomolecular functions and biomedicine. The molecular mechanism of protein-ligand binding remains an active research topic, despite of  enormous effort in the past few decades. The present work qualitatively demonstrates that   protein-ligand binding gives rise to protein rigidity enhancement.    Further   study shows that rigidity index alone is able to quantitatively describe protein-ligand binding affinities. The fact that rigidity index based scoring function,  RI-Score, offers the best binding affinity prediction over two benchmark data sets indicates that rigidity strengthening is a dominant mechanism in protein-ligand binding. Therefore, protein flexibility reduction drives protein-ligand binding. It is well known that protein flexibility is associated with many protein functions. The present finding strongly suggests that agonist protein receptor binding inhibits protein functions through protein flexibility reduction. Additionally, a significant correlation of the nearest four layers of residues to protein-ligand binding affinities is discovered, which has a nontrivial ramification to drug and protein design. 

{\bfseries{Supporting Information Available}} \url{SI\_RI-score.pdf} provides discussion of dataset preparation, and some additional results for PDBBind v2007 core set and PDBBind v2013 core set.

\section*{Acknowledgments}

This work was supported in part by NSF grants  IIS-1302285   and DMS-1160352 and
MSU Center for Mathematical Molecular Biosciences Initiative.

\bibliographystyle{ieeetr}
\bibliography{refs}

\end{document}